\documentclass[lettersize,journal]{IEEEtran}

\usepackage{blindtext, graphicx}
\usepackage{csquotes}
\usepackage{cite}
\usepackage{amsthm}
\ifCLASSINFOpdf
\else
\fi
\usepackage[cmex10]{amsmath}
\usepackage{optidef}
\usepackage{array}
\usepackage[tight,footnotesize]{subfigure}
\usepackage{stfloats}
\usepackage{float}
\usepackage{diagbox}
\usepackage{soul}

\def\*#1{\mathbf{#1}}

\usepackage{xcolor}

\usepackage{amsfonts}

\begin{document}

\title{How to Identify and Authenticate Users in \\Massive Unsourced Random Access}

\author{Rados\l{}aw~Kotaba,~\IEEEmembership{Member,~IEEE},
        Anders E. Kal{\o}r,~\IEEEmembership{Student Member,~IEEE},
        Petar~Popovski,~\IEEEmembership{Fellow,~IEEE},\\
        Israel Leyva-Mayorga,~\IEEEmembership{Member,~IEEE},
        Beatriz Soret,~\IEEEmembership{Member,~IEEE},
        Maxime~Guillaud,~\IEEEmembership{Senior Member,~IEEE}, and
        Luis G. Ord\'o\~nez,~\IEEEmembership{Member,~IEEE}

\thanks{R. Kotaba, A. E. Kal{\o}r, P. Popovski, I. Leyva-Mayorga, and B. Soret are with the Department of Electronic systems, Aalborg University, Aalborg 9220, Denmark  (e-mail:
\{rak, aek, petarp, ilm, bsa\}@es.aau.dk).}
\thanks{M. Guillaud, L. G. Ord\'o\~nez are with Huawei Technologies, Paris, France (e-mail:
maxime.guillaud@huawei.com, luis.ordonez@huawei.com).}
}



\maketitle

\begin{abstract}
Identification and authentication are two essential features for traditional random access protocols. In ALOHA-based random access, the packets usually include a field with a unique user address. However, when the number of users is massive and relatively small packets are transmitted, the overhead of including such field becomes restrictive. In unsourced random access (U-RA), the packets do not include any address field for the user, which maximizes the number of useful bits that are transmitted. However, by definition an U-RA protocol does not provide user identification. This paper presents a scheme that builds upon an underlying U-RA protocol and solves the problem of user identification and authentication. In our scheme, the users generate a message authentication code (MAC) that provides these functionalities without violating the main principle of unsourced random access: the selection of codewords from a common codebook is i.i.d. among all users.
\end{abstract}


\IEEEpeerreviewmaketitle

\section{Introduction}\label{sec:intro}
One of the hallmarks of the fifth generation (5G) wireless systems and beyond is massive Internet of Things (IoT) connectivity~\cite{bockelmann2018towards}. A scenario for massive IoT access features a large number of devices (typically in the order of thousands) connected to a Base Station (BS), each being sporadically active and sending short data packets (e.g., a few kilobytes or bytes). This sporadic activation entails that the set of devices trying to access at a given instant is unknown, thereby requiring random access protocols.

In the classical ALOHA model for random access~\cite{abramson1970aloha}, a packet is the smallest, atomic unit of information. The analyses in massive access scenarios are usually performed with an infinite population, where the number of users is $N\rightarrow\infty$. However, in order to examine the fundamental performance bounds of massive access protocols, one needs to look into the structure of the packet. This is where the assumption $N \rightarrow \infty$ leads to a paradox: to make user identification possible, a field with a unique user address of $\approx \log_2N$ bits must be included in a packet of finite and relatively short length. To deal with this paradox, two information-theoretic approaches have been introduced.
In the many access channel~\cite{chen2017capacity} the number of users is given as a function of the codeword length, which allows to preserve identification capabilities even as both tend to infinity.

Differently from this,~\cite{unsourced_polyanskiy} addresses the problem of $N\rightarrow\infty$ with finite blocklength (FBL) packets by assuming that a packet does not contain the address of the sender. This makes the access scheme \emph{unsourced}, and leads to the case in which all users share the same codebook. 
While U-RA was initially proposed as a theoretically elegant scheme, it can also be justified by the desire to simplify the receiver and reduce the communication overhead. This is particularly important for short IoT packets where the address field can constitute a large portion of the packet\cite{shortpacketsurvey}.

The unsourced, uncoordinated nature of the problem and the FBL effects have implications in the design of practical low-complexity coding schemes, which has been the focus of several works. Bounds of the performance of finite-length codes were derived in the initial paper by Polyanskiy~\cite{unsourced_polyanskiy}, and later generalized to the quasi-static fading channel~\cite{Kowshik2019}. The basic unsourced random access was extended to the case with a large number of antennas in~\cite{Fengler2019}, and the impact of correlated activations was studied in~\cite{Stern2019}. 

Despite its benefits in terms of efficiency, U-RA keeps the question of user identification (and, consequently, user authentication) open. In this paper, we aim to answer the following: \emph{assuming that a given protocol for unsourced random access is available as a black box, how can it be extended to support user identification and authentication?} 
Rather than deferring this question to the higher layers or additional transmissions, in this contribution we present a scheme that enables those functionalities at the lower layers, in a way that is consistent with the paradigm of U-RA, i.e., when users share the same codebook. 
In that sense, the main contribution of our scheme is that it enables the identification and authentication of users over U-RA; the potential performance gains compared to sourced random access is of secondary importance.

\begin{figure}[t]%
\centering
\subfigure[Downlink phase.]{
\label{fig:system:downlink}%
\includegraphics[width=0.4\textwidth]{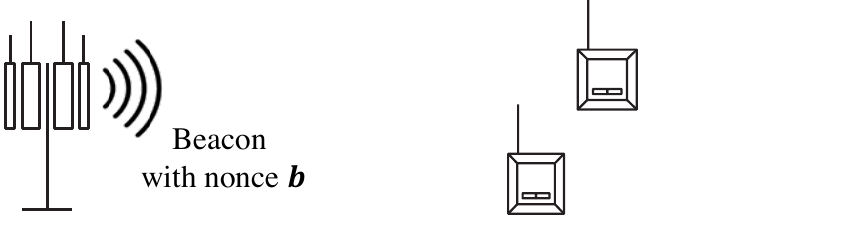}}%
\hfill
\subfigure[Uplink phase.]{
\label{fig:system_uplink}%
\includegraphics[width=0.4\textwidth]{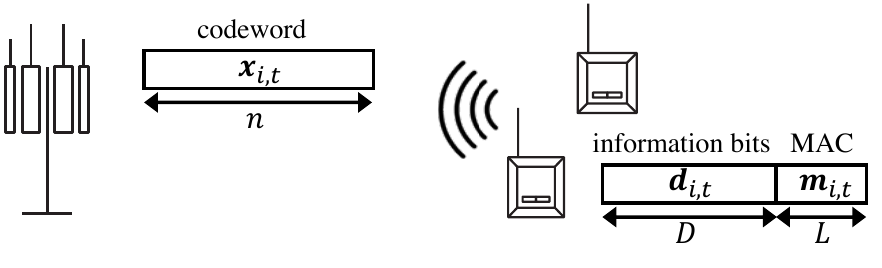}}%
\caption{The two-step procedure.}
\label{fig:system_model}
\end{figure}

The key idea is to generate and append a message authentication code (MAC)\footnote{To avoid confusion between this term and the widely-used acronym for medium access control, the latter is avoided throughout the paper.} to the packets (rather than an explicit address), which enables the identification and authentication of the users while complying with the main assumptions of U-RA. 
For this, we employ a two-step procedure as illustrated in Fig.~\ref{fig:system_model}. First, the BS broadcasts a beacon with a \emph{nonce} to the users prior to data transmission. A nonce is an arbitrary number generated periodically by the BS that is allowed to be used only once by each node to prevent replay of messages. Then, each active user generates a MAC based on the nonce, a secret key known only by the user and the BS (e.g., pre-shared using Universal Subscriber Identity Module (USIM) as in LTE~\cite{etsits133501}), and the data to be sent; this field is appended to the packet and transmitted as shown in Fig.~\ref{fig:system_uplink}.

\section{System model}
\label{sec:ura}
We study the massive random access scenario as described by Polyanskiy~\cite{unsourced_polyanskiy}, where $N\rightarrow\infty$ users communicate through a time-slotted channel with a single BS. Although the proposed scheme works without modifications with a (potentially massive) MIMO BS, we assume a single antenna BS to simplify the presentation. At each time slot, $K$ out of the $N$ users are active and send messages $\mathcal{W}= W_1,W_2,\ldots,W_K$ in the uplink, where $W_i$ is drawn independently and uniformly at random from the message set $\mathcal{M}=\{1,2,\ldots,M\}$. For typical massive IoT scenarios, $K$ will be in the range of $50$ to a few hundreds~\cite{decurninge21}. All users share the same encoder $f:[M]\to \mathcal{X}^n$, and use it to construct the codewords $\*x_1,\*x_2,\ldots,\*x_K$ as $\*x_i=f(W_i)$, which are subject to the constraint $\|\*x_i\|^2_2\le nP$, where $P$ is the average energy per symbol.
The codewords are transmitted over a permutation-invariant and memoryless multiple access channel $P_{Y|X_1^K}:\mathcal{X}^{n \times K}\to\mathcal{Y}^n$, i.e., it satisfies $P_{Y|X_1^K}(\*y|\*x_1,\ldots,\*x_K)=P_{Y|X_1^K}(\*y|\*x_{\pi(1)},\ldots,\*x_{\pi(K)})$ for any $\*y\in\mathcal{Y}^n$ and $\*x_1,\ldots,\*x_K\in\mathcal{X}^{n\times K}$, and any permutation $\pi$.

We assume that the BS periodically broadcasts a beacon in the downlink as depicted in Fig.~\ref{fig:system_model}. The beacon includes the necessary information for the users to synchronize, to obtain the main configuration parameters, and to estimate and invert the channel. 
To keep the presentation simple and aligned with~\cite{unsourced_polyanskiy}, we assume that channel inversion is perfect, so that fading can be neglected\footnote{We note that the users who cannot perform inversion due to poor channel conditions can simply remain inactive, which leads to the problem that is structurally the same.} and the uplink transmissions are only affected by additive white Gaussian noise, denoted by $\*z\sim \mathcal{N}(\*0,\sigma \*I_n)$. Consequently, the resulting Gaussian multiple access channel model at a given time slot is 
\begin{equation}
\*y = \sum_{i=1}^{K} \*x_i + \*z
\end{equation}

At the BS, the decoder $g:\mathcal{Y}^n\to [M]^K$ outputs an unordered list of $K$ messages from $\mathcal{M}$. In line with the U-RA literature~\cite{unsourced_polyanskiy}, we assume that $K$ is fixed and known to the decoder. 
We note that this assumption allows the codebook to be designed based on $K$, which does not reflect a true random access scenario. In practice, the codebook would have to be designed based on the expected maximum (or average) number of active users instead. Similarly, in practical implementations the decoder, rather than outputting a fixed number of messages, might rely on separate activity detection~\cite{decurninge21}.

An error occurs whenever the $g(\*y)$ does not contain a transmitted message, or if multiple users transmit the same message. More specifically, an error for user $i$ is defined as $E_i=\{W_i\notin g(\*y)\}\cup\{W_i=W_j\text{ for some } j \neq i\}$. Note that since we assume the decoder always outputs $K$ messages, it implies that for each error $E_i$, the list $g(\*y)$ must contain a message which was not transmitted by any of the devices. We shall refer to this set 
$ g(\*y) \setminus \mathcal{W} $
as decoder false positives. Denoting by $k_{\text{TP}}$ the number of genuine (true positive) messages and by $k_{\text{FP}}$ the number of false positives in the set $g(\*y)$, we have that $k_{\text{TP}}+k_{\text{FP}}=K$.

\section{Identification and authentication in unsourced random access protocols}

The key idea behind the proposed scheme is to generate MAC that enables identification and authentication of the users and that can be applied to U-RA protocols. The MAC $\*m_i=\{0,1\}^L$ is generated by user $i$ based on its data $\*d_i\in \{0,1\}^D$ of size $D$, its secret key $\*k_i$, and a nonce $\*b$. The secret key is fixed and only known by the corresponding user and the BS, e.g., pre-shared using USIM as in LTE~\cite{etsits133501}. The MAC length $L$ is fixed and independent from the other parameters. 

\begin{figure}[t!]	
	\centering
	\includegraphics[width=1.0\linewidth]{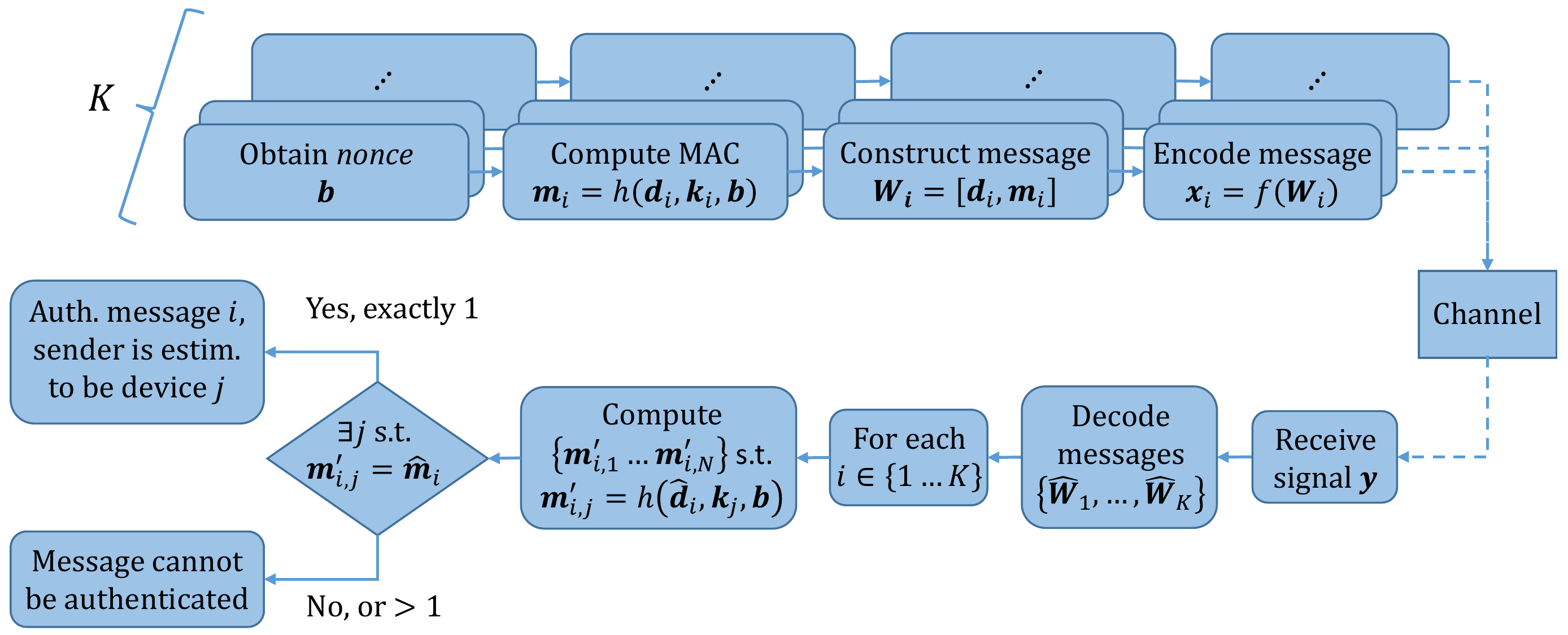}
	\caption{Block diagram of the proposed scheme including message generation and subsequent decoding, authentication and identification. 
	}
	\label{fig:algo}
\end{figure}

Our scheme is divided into phases as shown in Fig.~\ref{fig:system_model}. At the beginning of each round, the BS generates a nonce and broadcasts it to all the devices. The nonce is a sequence or pseudo-random number that changes in each round but is otherwise public. Once the nonce is received, a given user $i$ generates the MAC $\*m_i$ based on the data bits it wants to transmit $\*d_i$, its secret key $\*k_i$, and the nonce $\*b$, i.e. $\*m_i=h(\*d_i,\*k_i, \*b)$, where $h(\cdot)$ is designed to be computationally hard to invert and have low collision probability (i.e., the output is approximately uniform for any input distribution). The user appends the MAC to the data to create a packet and transmits it, as shown in Fig.~\ref{fig:system_uplink}.
At the BS, the packets are first decoded to extract $[\widehat{\*d}_i,\widehat{\*m}_i]$ tuples.
For each, the message authenticity can be verified and the identity of the sender determined by computing the MACs of the data part $h(\widehat{\*d}_i,\*k_j,\*b)$ with different secret keys $\*k_j$ and comparing them with the MAC in the received packet $\widehat{\*m}_i$. If a match is found, the authenticator declares the user with the matching key to be the potential transmitter. The full scheme is depicted in Fig.~\ref{fig:algo}.

While a nonce is commonly used to prevent replay attacks, in our scheme it has the additional function of randomizing the MAC. That is, without a nonce, a particular piece of data and secret key from a given device would always produce the same MAC, which violates the assumption that all codewords are equally likely. 
Typical methods to generate the MAC include, e.g., symmetric key cryptography as in AES-CMAC (RFC 4493), used in LoRaWAN, or a HMAC (RFC 2104). Any of these methods can be applied to our scheme, so the MAC is computationally challenging to guess without the secret key.

Note that in our scheme \emph{cryptographic errors}, which we define as any instance where the matching MAC is generated by a key that does not belong to the actual sender, can occur. 
They are possible since: 1) the generated MAC might not be a unique identifier for the user (unlike the actual address) and 2) the BS must generate many MACs with different secret keys to find the one that matches the one in the received packet.

Therefore, several tradeoffs arise. The first one is between the length of the metadata and the amount of \emph{cryptographic errors}, where in the extreme case with no metadata (i.e. neither MAC nor address) identification and authentication cannot be provided.
Meanwhile, longer packets entail higher energy.
Another tradeoff involves the computational complexity and probability of cryptographic errors that both increase with the number of devices supported by the system\footnote{It could be argued that the scheme is not practical as $N\rightarrow\infty$. However, in practice good performance was observed for $N$ as large as $10^6$ and $K>100$.}.

\section{Cryptographic errors: collisions, false positives and misidentifications}

The probability of decoder false positives describes only the physical layer performance of U-RA.
The full characterization of the proposed scheme has to take into account also potential cryptographic errors, erroneous acceptance of false positives, and misidentification events. For the purpose of this evaluation, we assume ideal MACs that are uniformly distributed, i.e., the probability that a given $(\textit{data}, \textit{key}, \textit{nonce})$ tuple produces a specific MAC of length $L$ is $p=2^{-L}$.

\subsection{Exhaustive search}
We first consider authentication using exhaustive search, where all keys are tried on each message. We start by studying the per-user cryptographic error probabilities. A genuine message $W'$ with data $\mathbf{d}'$ transmitted by user $u'$ will fail to be authenticated whenever any of the keys from users $u\neq u'$ produces the same MAC when applied to $\mathbf{d}'$. We refer to those events as \emph{type 1} errors.
Since there are $N-1$ other keys, the type 1 event happens with probability
\begin{equation}
    p_{\text{t1}} = 1-(1-p)^{N-1}
\end{equation}
Because we assume that each user transmits at most one message per round, an error occurs also when the key of user $u'$ produces a valid MAC for any of the other decoded messages in 
$g(\mathbf{y}) \setminus \{W'\}$. 
Given that there are $K-1$ other decoded messages, this \emph{type 2} error happens with probability
\begin{equation}
    p_{\text{t2}} = 1-(1-p)^{K-1}.
\end{equation}
Taking into account both types of errors, the probability that a genuine message is successfully authenticated is
\begin{equation}
    p_{\text{s\_auth}} = (1-p)^{N+K-2}.
    \label{eq:ex:p_e_auth}
\end{equation}

Another type of event is when a false positive message produced by the decoder is erroneously authenticated. While \eqref{eq:ex:p_e_auth} is conditional on the fact that there is at least one key that authenticates the message, here we cannot assume that. Since the keys from the genuine messages cannot be used again without causing type 2 error, there are $N-k_{\text{TP}}$ keys that can potentially decode the false positive message without resulting in a collision. Since each of these keys accepts the message with probability $p_{\text{s\_auth}}$, the probability of accepting a false positive message from the decoder is
\begin{equation}
\begin{split}
    p_{\text{fp\_auth}} &= (N-k_{\text{TP}}) \,p\, p_{\text{s\_auth}}\\
    &=(N-k_{\text{TP}})p(1-p)^{N+K-2}.
\end{split}
\label{eq:ex:fp_auth}
\end{equation}
Note that the authenticator is generally unable to determine whether a message that \emph{fails} to be authenticated belongs to the set of decoder true positive or decoder false positive messages. The only exception to this is the special case in which no key is able to decode a given message, which can only happen for false positive messages. The probability that this happens for a given false positive message is $p_{\text{d\_fp}}=(1-p)^N$.

\subsection{Heuristic search}
We now turn our attention to the heuristic search, in which the authenticator tries keys only until it finds a matching key. While more efficient, this approach cannot detect type 1 and type 2 errors defined above, and thus the probability of erroneously authenticating a message increases. 

Providing exact analytical expressions for the heuristic case proves to be difficult, due to the dependency on the order in which packets are authenticated, the number of decoder false positives and true positives, and how they are interleaved. To that end, we will provide only approximations, noting that they are very close to the true values. 
We shall assume without loss of generality that the decoded messages are authenticated in the order $\widehat{W}_1, \widehat{W}_2, \ldots, \widehat{W}_K$. Furthermore, we will neglect the events where the sender of message $\widehat{W}_j$ becomes incorrectly identified as the sender of one of the previous messages $\widehat{W}_1,\ldots,\widehat{W}_{j-1}$, which happens with very low probability\footnote{Note that we do not neglect misidentification events in general, but only the case where specific user authenticates a specific message, which is tied to the probability $p$ and hence very low.}.

We first consider the probability of correctly authenticating a genuine message. In the heuristic search case the successful authentication of message $W_j$ can happen even if there are cryptographic collisions, as long as the correct user happens to be tested first. For a set of $i$ successfully authenticating keys, this happens with probability $1/i$. By marginalizing over the number of keys \emph{additional} to the genuine key we obtain
\begin{equation}
    p_{\text{s\_auth},j} = \sum_{i=0}^{N_j-1} \binom{N_j-1}{i} p^i (1-p)^{N_j-1-i} \left(\frac{1}{1+i}\right),    \label{eq:he:p_succ}
\end{equation}
where $N_j$ is the number of remaining keys which is the total number of keys, $N$, minus those that have authenticated any of the previous messages. $N_j$ is nonincreasing, and $N_j\ge N-j+1$ since the authenticator may have been unable to authenticate some of the previous $j-1$ messages. As already mentioned, in the heuristic approach the detection of collisions (type 1 and type 2 errors) is not possible, which can result in misidentification, i.e., attributing a genuine message to the wrong user. The probability of misidentifying the $j$-th message is the probability that one or more of the $N_j-1$ non-genuine keys authenticate the message before the correct one:
\begin{equation}
    \begin{split}
    p_{\text{mis\_id},j} &= \sum_{i=1}^{N_j-1} \binom{N_j-1}{i} p^i (1-p)^{N_j-1-i} \left(1-\frac{1}{1+i} \right)\\
    &= 1-p_{\text{s\_auth},j}.
    \label{eq:he:misid}
    \end{split}
\end{equation}

On the other hand, if the message is a false positive, the probability of accepting it is equal to the probability of having at least one key which produces a matching MAC:
\begin{equation}
    p_{\text{fp\_auth},j} = 1 - (1-p)^{N_j}.
    \label{eq:he:fp_auth}
\end{equation}

We note that from the point of view of the receiver there is no difference between misidentification and false positive authentication, hence, the total error probability should include both. For a given packet, which is genuine with probability $p_{\text{TP}}$ and a false positive with probability $p_{\text{FP}}$ we obtain
\begin{equation}
    p_{\text{mis\_auth},j} = p_{\text{TP}}  p_{\text{mis\_id},j} + p_{\text{FP}} p_{\text{fp\_auth},j}.
    \label{eq:he:mis_auth}
\end{equation}

Lastly, let us remark that when $N\gg K$, we have that $N_j \approx N$. 
By making this substitution in  \eqref{eq:he:p_succ}~-~\eqref{eq:he:mis_auth},
we obtain a rigorous upper bound on the probability of each type of error.
Furthermore, they become independent of packet number and allow us to drop the subscript $j$ which simplifies the comparison between the exhaustive and heuristic approach.

\begin{figure}[t!]	
	\centering
	\includegraphics[width=1.0\linewidth]{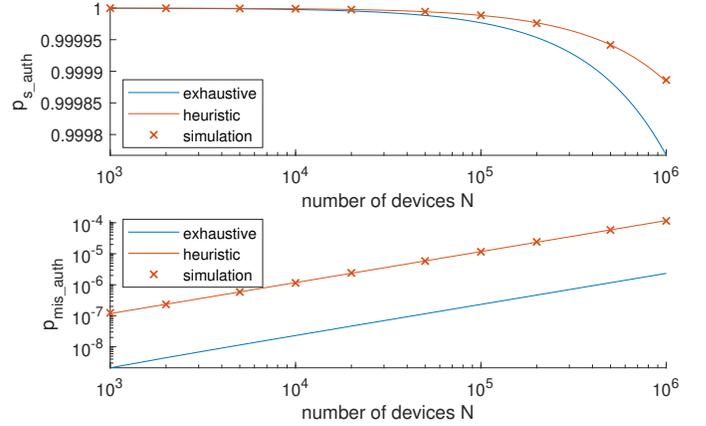}
	\caption{Probability of successful authentication and probability of mis-authentication as a function of the total number of devices $N$. The number of messages is $K=100$, $p_{TP}=0.99$, $p_{FP}=0.01$, MAC length $L=32 \text{bits}$.}
	\label{fig:mac_perf}
\end{figure}

In Fig. \ref{fig:mac_perf} we show the probability of successful authentication and probability of mis-authentication as a function of the total number of devices $N$ for the exhaustive and heuristic search. 
In addition to the small gain in terms of success probability, the latter method allows to reduce the complexity as, on average, it requires only half of the MAC checks (assuming the probability of transmission is uniform across the devices). This is at the cost of an increased probability of mis-authentication.
Since the eq. \eqref{eq:he:p_succ} and \eqref{eq:he:mis_auth} used to produce the solid red curves are approximations that neglect some of the effects mentioned earlier, we provide also the results obtained through numerical simulations. Clearly, the differences are very minor making the approximations a viable tool.

\subsection{Spoofing attacks}
It is of interest to consider what happens when an attacker sends a forged message with the intent of getting it accepted by the authenticator. Without private keys the attacker is not able to compute the correct MAC for the spoofed $\textit{data}$ and current $\textit{nonce}$ so it has to generate MAC bits at random.
As such, from the cryptographic point of view, the message acts as a false positive, and the transmitter cannot target a specific device (i.e. it cannot choose whom it is impersonating).
However, from the physical layer point of view this is an actual transmitted codeword, and as such subject to probability of decoding $p_\mathrm{TP}$, so the total probability of successful spoof is
\begin{equation}
    p_{\mathrm{s\_spoof}}=p_{\mathrm{TP}}\left(1-(1-p)^N\right).
\end{equation}
This is to be compared to the traditional frame structure where the source address is included in the packet. In that case, the authenticator only tries the single MAC associated to that user, and the spoof attack is successful with probability $p_{\mathrm{TP}}p$.

\section{Results}\label{sec:res}
\begin{figure}[t!]	
	\centering
	\includegraphics[width=1.0\linewidth]{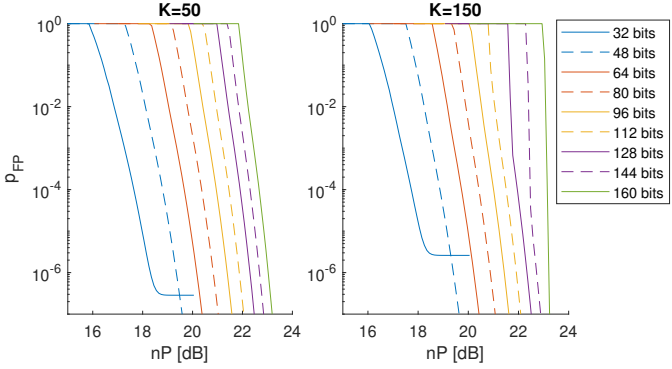}
	\caption{Achievable physical layer error probability as a function of the energy per codeword and packet size $B$. Two different values of $K$ are considered.}
	\label{fig:snr_vs_outage}
\end{figure}

\begin{figure}[t!]	
	\centering
	\includegraphics[width=0.86\linewidth]{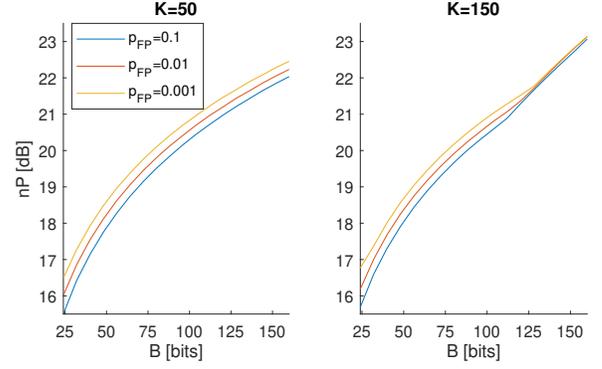}
	\caption{Minimum energy required to achieve fixed physical layer error probability as a function of the number of information bits $B$.}
	\label{fig:rate_vs_snr}
\end{figure}

We start by looking into the physical layer performance. The results were obtained based on the random coding bound given in \cite[eq. (3)-(10)]{unsourced_polyanskiy}. The codeword length (number of symbols) was chosen to be $n=2^{15}=32768$.
In Fig.~\ref{fig:snr_vs_outage} we depict the achievable error probability as a function of the energy per codeword $nP$. The values are shown for a range of packet sizes $B$ and for $K=50$ and $K=150$.
In line with the assumption that two users selecting the same message is considered an error (c.f. Section~\ref{sec:ura}), each curve has a floor at $\binom{K}{2}/M=\binom{K}{2}/2^B$ (visible only for $32$ bits).
In general, the higher B and K are, the steeper the curves become and the transition from almost certain error $p_\mathrm{FP}\approx 1$ to very high reliability (such as $p_\mathrm{FP}<10^5$) becomes increasingly abrupt.
This is even better explained with Fig. \ref{fig:rate_vs_snr} which shows the energy per codeword as a function of $B$ for fixed error rates. Firstly, as the packet size increases, less energy is needed to decrease $p_\mathrm{FP}$. For example, with $K=50$, when $B=25\, \mathrm{b}$ improving error rate from $10^{-1}$ to $10^{-3}$ requires $1\, \mathrm{dB}$, while with $B>100\, \mathrm{b}$ the same shift requires less than $0.5\, \mathrm{dB}$. 
Secondly, there is a point where the system turns from being noise-limited to interference-limited (curves merging). Such a transition occurs for lower packet sizes the more simultaneous messages $K$ there are. 

\begin{figure}[t!]	
	\centering
	\includegraphics[width=0.865\linewidth]{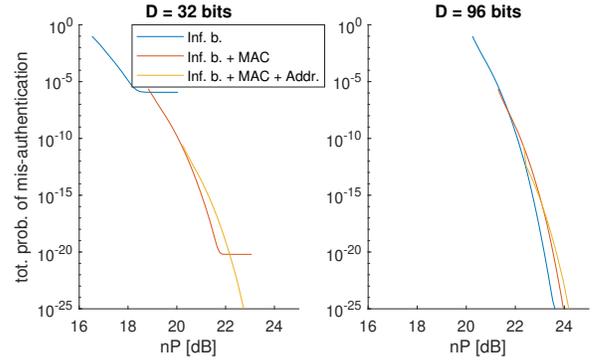}
	\caption{Total probability of mis-authentication as a function of energy, taking into account both physical and cryptographic layer. The number of messages is $K=100$ and $N=10^5$.}
	\label{fig:tot_misauth}
\end{figure}

In Fig. \ref{fig:tot_misauth} we combine all the earlier insights and look into the total probability of mis-authentication that takes into account both the physical and cryptographic layer performance. These results are obtained assuming a population of $N=10^5$ users and $K=100$ messages.
In the plots, the blue line represents a packet consisting solely of the information bits, i.e. $B=D$. Since there is no additional means of authentication, every decoded packet is accepted and hence $p_{\text{miss\_auth}}=p_{\mathrm{FP}}$. 
The red line denotes our proposed scheme in which the packet consists of information bits and a MAC, that is, in total $B=D+L$, where $L=32\, \mathrm{b}$ is fixed. The values reported here correspond to the exhaustive search variant, hence, the total probability of mis-authentication is $p_{\text{miss\_auth}}=p_{\mathrm{FP}}p_{\text{fp\_auth}}$ with 
$p_{\text{fp\_auth}}$
given by \eqref{eq:ex:fp_auth}. 
Lastly, the yellow curve represents the classic packet structure, where the address is also included (here $A=32\, \mathrm{b}$ as well) which yields $B=D+L+A$. In such case, the receiver checks only one key corresponding to the given address, hence we have that $p_{\text{miss\_auth}}=p_{\mathrm{FP}}p$.
It is important to keep in mind that the most basic mode of operation (blue) does not provide any way of identifying the users,
and as such it is not directly comparable with the other two.
Furthermore, it might not provide sufficient level of reliability when the packets are very short, which is due to the floor on $p_{\mathrm{FP}}$.
What might be surprising, is that the classic packet structure actually performs slightly worse than our proposed scheme (at least until $10^{-20}$ level which should be more than enough). This is because even though the probability of MAC collision is significantly lower, the packet needs to be larger to accommodate the address, which requires higher energy.

\section{Conclusions}\label{sec:concl}
In this paper we proposed a method to introduce identification and authentication capabilities to algorithms that follow the framework of unsourced random access.
Our scheme adds very limited amount of metadata to the communication, which is especially important for short IoT packets. 
Furthermore, as a consequence of not including explicit user identification, the packets are fully anonymous to everyone except the BS, which opens the door to new use cases and applications. This is in contrast to traditional protocols, where only the message content is assumed to be secret while the identities are public.
The extra functionalities come at the cost of increased processing at the receiver.
However, our results show that by avoiding the address we can simultaneously improve the spectral efficiency, and, for a given given energy per codeword, decrease the overall mis-authentication probability compared to the case where the address is included in the packet.

\appendices

\ifCLASSOPTIONcaptionsoff
  \newpage
\fi

\bibliographystyle{IEEEtran}
\bibliography{Biblio}

\end{document}